\documentclass[twocolumn]{aastex63}
\usepackage{hyperref}
\usepackage{bm}
\usepackage{multirow}
\usepackage{gensymb}
\usepackage{enumitem}
\usepackage{tabularx}
\usepackage[flushleft]{threeparttable}
\usepackage[]{color}

\usepackage{natbib}
\usepackage{float}

\usepackage{amsfonts,amsmath,amssymb}
\usepackage{graphicx}
\usepackage{color}
\usepackage{xcolor}
\usepackage{soul}
\usepackage{url}
\usepackage{float}

\newcommand{\xstar}{\textsc{\scriptsize XSTAR }}

\newcommand{\athena}{\textsc{Athena\scriptsize ++ }}

\renewcommand{\vec}[1]{\mathbf{#1}}
\newcommand{\e}[1]{\times10^{#1}}
\newcommand{\E}[1]{\left\langle#1\right\rangle}
\newcommand{\uvec}[1]{\hat{\boldsymbol{#1}}}
\newcommand{\W}[0]{\textsc{\tiny W}}
\newcommand{\sobeq}[0]{\mathrel{\stackrel{\makebox[0pt]{\mbox{\normalfont\tiny Sob}}}{=}}}

\DeclareMathAlphabet\mathbfcal{OMS}{cmsy}{b}{n}

\defcitealias{Proga99}{PSD99}
\defcitealias{Proga98}{PSD98}
\defcitealias{CAK}{CAK}
\defcitealias{OCR}{OCR}
\defcitealias{PSK}{PSK}
\defcitealias{Dannen19}{D19}
\defcitealias{Dannen20}{D20}
\defcitealias{Waters21}{WPD21}
\defcitealias{Begelman83}{BMS83}

\def\CIVdbl{{\rm C~}\kern 0.1em{\sc iv}~$\lambda\lambda 1548, 1550$} 
\def\OVIIIi{\hbox{{\rm O}\kern 0.1em{\sc viii}}}
\def\SiXIVi{\hbox{{\rm Si}\kern 0.1em{\sc viii}}}
\def\OVIII{\hbox{{\rm O}\kern 0.1em{\sc viii}~{\rm Ly}$\alpha$}}
\def\SiXIV{\hbox{{\rm Si}\kern 0.1em{\sc viii}~{\rm Ly}$\beta$}}
\def\FeXXV{\hbox{{\rm Fe}\kern 0.1em{\sc xxv}}}
\def\FeXXVI{\hbox{{\rm Fe}\kern 0.1em{\sc xxvi}}}
\def\FeXXVK{\hbox{{\rm Fe}\kern 0.1em{\sc xxv}~{\rm K}$\alpha$}}
\def\FeXXVIK{\hbox{{\rm Fe}\kern 0.1em{\sc xxvi}~{\rm K}$\alpha$}}

\newcommand{\beq}{\begin{equation}}
\newcommand{\seq}{\end{equation}}

\renewcommand{\v}[1]{\ensuremath{\mathbf{#1}}} 
\renewcommand{\d}[2]{\frac{d #1}{d #2}} 

\newcommand{\f}{\frac}  
\renewcommand{\exp}[1]{\exp\left( #1 \right)} 

\usepackage{letltxmacro,amsmath}
\LetLtxMacro{\originaleqref}{\eqref}
\renewcommand{\eqref}{Eq.~\originaleqref}

\submitjournal{ApJ}
\shorttitle{Disk Radiation Fields}
\shortauthors{Smith, Proga, Dannen, Dyda, \& Waters}
\begin{document}
\title{Position dependent radiation fields near accretion disks}

\correspondingauthor{Kara Smith}
\email{smith99@unlv.nevada.edu}
\author[0000-0001-5506-1968]{Kara Smith}
\affiliation{Department of Physics \& Astronomy \\
University of Nevada, Las Vegas \\
4505 S. Maryland Pkwy \\
Las Vegas, NV, 89154-4002, USA}
\affiliation{Nevada Center for Astrophysics \\
University of Nevada, Las Vegas \\
4505 S. Maryland Pkwy. \\
Las Vegas, NV 89154, USA}
\author[0000-0002-6336-5125]{Daniel Proga}
\affiliation{Department of Physics \& Astronomy \\
University of Nevada, Las Vegas \\
4505 S. Maryland Pkwy \\
Las Vegas, NV, 89154-4002, USA}
\affiliation{Nevada Center for Astrophysics \\
University of Nevada, Las Vegas \\
4505 S. Maryland Pkwy. \\
Las Vegas, NV 89154, USA}
\author[0000-0002-5160-8716]{Randall Dannen}
\affiliation{Department of Physics \& Astronomy \\
University of Nevada, Las Vegas \\
4505 S. Maryland Pkwy \\
Las Vegas, NV, 89154-4002, USA}
\affiliation{Nevada Center for Astrophysics \\
University of Nevada, Las Vegas \\
4505 S. Maryland Pkwy. \\
Las Vegas, NV 89154, USA}
\author[0000-0002-1954-8864]{Sergei Dyda}
\affiliation{Department of Astronomy \\
University of Virginia \\
530 McCormick Rd \\
Charlottesville, VA 22904, USA}
\author[0000-0002-5205-9472]{Tim Waters}
\affiliation{Theoretical Division, Los Alamos National Laboratory, \\
NM 87545, USA}

\begin{abstract}

In disk wind models for active galactic nuclei (AGN) outflows, high-energy radiation
poses a significant problem wherein the gas can become overionized, 
effectively disabling what is often inferred to be the largest force acting on the gas:
the radiation force due to spectral line opacity.
Calculations of this radiation force depend on the magnitude of ionizing radiation, 
which can strongly depend on the position above a disk where the radiation is anisotropic. 
As our first step to quantify the position and direction dependence of the radiation field, 
we assumed free streaming of photons and computed energy distributions of the mean intensity 
and components of flux as well as energy-integrated quantities such as mean photon energy. 
We find a significant dependence of radiation field properties on position, 
but this dependence is not necessarily the same for different field quantities.
A key example is that the mean intensity is much softer than the radial flux at many points near the disk.
Because the mean intensity largely controls ionization, this softening decreases 
the severity of the overionization problem.
The position dependence of mean intensity implies the position dependence of gas opacity, 
which we illustrate by computing the radiation force a fluid element feels in an accelerating wind.
We find that in a vertical accelerating flow, 
the force due to radiation is not parallel to the radiation flux. 
This misalignment is due to the force's geometric weighting 
by both the velocity field's directionality and the position dependence of the mean intensity.

\end{abstract}

\section{Introduction}\label{sec:intro}

The key properties of active galactic nuclei (AGN), such as compactness, 
high luminosity, time variability, spectral energy distribution (SED), 
and spectral features can be well-explained as consequences of disk accretion 
onto a supermassive black hole (SMBH). Spectral features such as broad emission lines (BELs) 
have for decades been used as a basis for classifying AGN, while broad absorption lines (BALs) 
in the ultraviolet (UV) are a spectacular manifestation 
for gas not only accreting onto an SMBH but also flowing outward.
Even though it has been shown the gas responsible for line emission/absorption 
is irradiated by the SMBH accretion disk 
\citep[e.g., see][for a textbook review]{Krolik99},
the structure and dynamics of the region where these broad lines 
are formed (the broad line region, BLR) remains elusive. 

Spectral features indicate that these accretion disks commonly produce outflows 
of matter in the form of jets and winds. The latter is one of the fundamental mechanisms 
by which the central SMBH interacts with its host galaxy 
\cite[e.g.,][and references therein]{Giustini19, Laha20}. 
Therefore, the gas outflows are important not only for the inner workings of AGN but also
for AGN feedback \citep{Fabian12}. The mechanisms launching 
and driving these accretion disk winds include 
thermal driving, magnetic driving, and radiation pressure due to photon 
scattering by spectral lines 
\citep[`line driving'; e.g., see for a review][and references therein]{Proga07}. 
While each of these three mechanisms requires its own methods of study 
and computation, a commonality between them is that each depends on the radiation 
fields of the accretion disk as the source of ionizing radiation. 

One major problem for any AGN wind model is overionization of the gas, 
which can both suppress or inhibit outflows and fail to reproduce observed line profiles, 
namely lines produced from C~{\sc iv}, Si~{\sc v}, and N~{\sc v}.
Even a clumpy magnetically driven outflow not shielded from the powerful AGN 
radiation will be overionized and not reproduce observed BALs \citep{dKB95}. 
In a line driving scenario, the problem is much more severe; an overionized gas cannot 
be accelerated by line driving at all \citep[e.g.,][]{sk90, PSK, Dannen19}.

Estimates of the ionizing radiation based on constraints of the SED 
and luminosity obtained by a distant observer show that for the so-called unobscured AGN, 
without shielding, the wind would be overionized \citep{Murray95}.
However, a fluid element in the wind is not a \textit{distant} observer; 
it sees the radiation source as a \textit{nearby} observer.
Regardless of which driving mechanism is favored or dominant, 
understanding the radiation fields seen by the fluid element in the wind 
is fundamental to determining if the wind is as overionized as the distant observer implies.

To understand the radiation fields as seen by the outflow, we compute mean intensity, 
flux, mean photon energy, and the line force as a function of position. 
We find that a simple multi-temperature blackbody treatment, which a distant observer sees, 
is woefully inadequate for describing the radiation very near to the disk.
We have developed a code capable of accurately computing radiation fields near the disk 
and can do so inexpensively in terms of computational resources.
In this paper, we use our code to explore variables that contribute to line driving.

The paper is organized as follows. In \S\ref{sec:prelim}, we give a basic overview 
of the theory behind radiation fields and line-driven winds from accretion disks.
In \S\ref{sec:calc}, we discuss the parametrization of the disk and the computation 
of radiation fields. In \S\ref{sec:res}, we discuss the results of the study;
including comparisons between SEDs at different points relative to the disk 
(\S\ref{sec:local vs distant}), the sensitivity of mean photon energy, $\E{h\nu}$ 
and mean ionizing intensity, $J_{\rm{ion}}$,
to changes in the coronal radiation (\S\ref{sec:x_sensitivity}), and finally, the differences 
in the direction between the radiation flux and the radiation force for a vertical velocity 
field (\S\ref{sec:misalign}).
The paper closes with a summary and discussion of the results in \S\ref{sec:conclusions}.

\section{Preliminaries}\label{sec:prelim}

\subsection{From Stars to Disks: Necessary Changes in Perspective.}\label{sec:star_v_disk}

Models of outflows in AGN are informed by the relatively well-understood winds from O stars 
\citep[for a textbook review see][and references therein]{Lamers-Cassinelli}. 
However, one must remember that stars are simpler than disks when discussing the radiation field. 
For example, stellar photospheres are relatively uniform in temperature; therefore, 
under free-streaming conditions, the specific intensity, $I_\nu$, above the star 
is direction independent inside the solid angle subtended  by the star, $\Omega_*$. 
Further, when free-streaming is applicable, $I_\nu$ equals the intensity 
of the stellar photosphere, $I_{\nu,0}$. 
Therefore, calculations that require integration over $\Omega_*$ can be simplified, e.g.,
when calculating the mean intensity, $J_\nu = (4\pi)^{-1}\oint_{\Omega_*} I_{\nu}\ d\Omega$, 
one can remove the term for specific intensity from the geometric integral, 
yielding the well-known relation $J_\nu = I_{\nu,0} \mathcal{W}(r)$, 
where $\mathcal{W}(r) = \frac{1}{2}(1 - \sqrt{1 - (r_*/r)^2})$ is the geometric dilution factor 
and $r_*$ is the stellar radius.
For $r \approx r_*$, $\mathcal{W}(r) \approx 1/2$ and for large distances, 
$\mathcal{W}(r) \approx (r_*/2r)^2$ \citep[see][]{Lamers-Cassinelli}.

Disks are flattened structures that are \textit{not} uniform in temperature.
Not only the distance, $r$ but also the viewing angle, $\theta$,
measured from the normal of the disk
are necessary variables to consider in describing any observed properties of the disk.
A nearby observer is taken to be a fluid element with a radius less than 
the disk’s outer radius, $r_{\rm out}$. For fluid elements very near to the disk, 
the mean intensity does approach $J_\nu \rightarrow \frac{1}{2} I_{\nu,0}$, 
as it does for stars.
However, the specific intensity above the disk is \textit{not} direction-independent 
within the solid angle subtended by the disk, $\Omega_D$.
Instead, the disk photospheric temperature and specific intensity are typically 
a decreasing function of the radius along the disk, $r_D$.
In a standard accretion disk theory \citep{Shakura-Sunyaev}, the disk photospheric temperature, 
$T(r_D)$, follows a $r_D^{-3/4}$ scaling for large $r_D$ (a similar scaling can also be derived 
for an irradiated disk)
Assuming that the disk radiates as a blackbody at every point, 
$I_{\nu,0}(r_D)$ follows the famous $r_D^{-3}$ scaling for large $r_D$, 
so near the disk, $J_\nu$ scales as $1/r^3$ rather than $1/r^2$.

If the disk were of constant specific intensity, then the radial radiation flux 
for a large $r$ and given $\theta$ would be $F_{\nu,r} = I_{\nu,0}A_D\cos\theta/r^2$, 
where $A_D$ is the area of the disk. 
The factor of $\cos\theta$ accounts for the geometrical foreshortening effect.
Because disks have non-uniform specific intensity profiles, we must integrate 
them over the entire disk surface.
A simple way to do this is to assume that the angular distance between points 
on the disk is negligible 
and only to integrate over radii \citep[see, e.g.,][]{Frank02}.
This is only a good approximation for the radial flux as seen by a very distant observer, 
though, so we denote it $F_{\nu,dist}$:

\begin{equation}\label{eq:F dist}
    F_{\nu,dist} = \f{2\pi\cos\theta}{r^2}\int_{r_{in}}^{r_{out}} I_{\nu, 0}(r_D)\cdot r_D\ dr_D,
\end{equation}
where $r_{in}$ is the radius of the disk's inner edge. 
This estimate of flux is not adequate for computing properties from the point 
of view of a nearby fluid element, as non-radial flux components comprise 
a significant portion of the total flux near the disk.
In fact, as we approach arbitrarily close to the disk, the radiation flux 
is dominated by local radiation.
Each point on the disk radiates a single-temperature blackbody,
not the multi-temperature blackbody produced by Equation~(\ref{eq:F dist}).
Even if a fluid element receives substantial radiation from the entire disk, 
assuming negligible angular differences between points on the disk can lead
to inaccurate estimates.
Finally, the scaling in Equation~(\ref{eq:F dist}) is $1/r^2$, but the flux should 
be proportional to $1/r^3$ near the disk, as noted above. 
Therefore, the distant observer approximation (Equation~\ref{eq:F dist}) 
cannot be applied to points close to the disk. However, it is still useful 
for benchmarking numerical calculations, as demonstrated by previous studies (e.g., see \citep[]{Proga98}).

\subsection{Line Driven Disk Winds}\label{sec:winds_theory}

One of the primary quantities derived from the radiation field 
is the radiation force due to lines, $\mathbfcal{F}^{rad}$. 
Winds can be launched when continuum photons interact with gas 
through scattering in spectral lines \citep[hereafter CAK]{CAK}. 
\citetalias{CAK} expresses the radial component of the radiation force, 
$\mathcal{F}^{rad}_r$, in terms of the radial radiation flux, $F_r$, as

\begin{equation}\label{eq:force mult def}
    \mathcal{F}^{rad}_r = M(t)\frac{\sigma_e}{c}F_r,
\end{equation}
where $M(t)$ is the force multiplier, which is 
a function of optical depth parameter $t$,
$\sigma_e$ is the electron scattering coefficient, and $c$ is the speed of light. 
\citetalias{CAK} defined $t$ using the Sobolev approximation \citep{Sobolev}, 
$t \equiv \sigma_e v_{th} \rho |dv_r/dr|^{-1}$, where $v_{th}$ is the thermal velocity 
of the gas, and $\rho$ is the density.
Notice that this definition of t assumes a radial flow; 
additionally, CAK's theory uses radial force and flux as analogs 
for their total values because the theory is developed for optically thin gas 
around a spherically symmetric star.
CAK’s theory has been successfully extended to a disk 
\citep[hereafter PSD98 and PSD99]{Proga98, Proga99}. 
However, the gas velocity is no longer radial, so instead.

\begin{equation}\label{eq:t def}
t \equiv \rho\sigma_e v_{th}\left|\frac{dv_\ell}{d\ell}\right|^{-1},
\end{equation}
where $\ell$ is the distance from the emitting point on the disk along 
a given line of sight defined by the unit vector $\uvec{n}$ and $v_\ell$ 
is the velocity along the line of sight: $v_\ell = \mathbf{v}\cdot\uvec{n}$.
To compute the radiation force, we must integrate the intensity 
over all such lines of sight, 
and because $dv_\ell/d\ell$ is direction dependent, $\mathcal{M}(t)$ 
cannot be removed from this integral. 
Thus, we have an expression for the radiation force,

\begin{equation}\label{eq:Frad_D}
    \mathbfcal{F}^{rad} = \oint_{\Omega_D} \mathcal{M}(t)\cdot\frac{\sigma_e I}{c}\uvec{n}d\Omega,
\end{equation}
where $I$ is the frequency integrated specific intensity of the disk. 
In a sense then, Equation~(\ref{eq:force mult def})
is the ``distant observer'' analog of Equation~(\ref{eq:Frad_D}), where $I$ is non-zero 
for only one direction: purely radial.
\citetalias{CAK} developed a correction to account for the changes in velocity 
along a non-radial line of sight. 
This correction, 
elaborated upon by \citet{friend86} and \citet{pauldrach86},
uses the symmetry of stars to integrate over all lines of sight analytically.
The line force remains purely radial, though, also due to symmetry.
For a disk, we cannot use such symmetry; 
instead, we must use fast numerical methods to account for sufficiently many lines 
of sight to the disk to sample $\Omega_D$.
One might assume that, generally, the line force has the same direction as the radiation flux 
-- as is done to derive Equation~(\ref{eq:force mult def}) --
even when that direction is not purely radial,
and that the biggest term of $dv_\ell/d\ell$ suffices to compute 
the underlying velocity shear tensor \citep{Proga98, Nomura20}.
In section \S\ref{sec:misalign}, however, we show that the force and flux 
are not aligned near a disk of non-uniform brightness.

The \citetalias{CAK} line-driven wind theory informed earlier studies 
of line-driven winds from disks around cataclysmic variables 
\citep[CVs;]{Proga98, Proga99} and in AGN \citep{Murray95, PK04}. 
In CVs, launching and driving winds are robust, 
as the winds are optically thin to continuum photons and the radiation 
is dominated by UV, similar to the O stars modeled by CAK. 
In AGN, however, the acceleration of a wind is less robust.
The wind can be accelerated to high velocities only if both 
the disk continuum radiation is strong in the UV \textit{and} 
the wind can be shielded from the ionizing radiation 
from the inner disk and the X-ray corona \citep{dyda23}.
Because AGN winds are much denser than those of O stars and CVs, 
gas launched from large radii can be accelerated when shielded 
by a dense flow launched from smaller radii that fail to escape 
the disk due to overionization \citep[hereafter PSK]{PSK}.

In \citetalias{PSK}, the radiation transfer of the ionizing radiation 
was treated in a simplified way, namely by accounting only 
for the attenuation of direct radiation. 
Such a treatment results in a lower limit for the mean intensity 
of ionizing radiation $J_{ion} \equiv \int_{\nu_H}^\infty J_\nu\ d\nu$,
where $\nu_H$ corresponds to 13.6 eV and $J_\nu$ is the mean intensity 
as a function of frequency. It does not account for scattered radiation found 
by \cite{Sim10} and \cite{Higginbottom14}.
The latter studies used Monte Carlo photoionization calculations 
and showed that radiation scattering can allow ionizing photons 
to propagate `around' the shielding inner flow and thereby increase 
the ionization of the outer outflow.
The techniques used by \cite{Sim10} and \cite{Higginbottom14} are computationally expensive, 
and as such, it is not yet feasible to use them in the context of radiation-hydrodynamical simulations,
wherein we would need to compute line distributions for each grid point at each time step.
It is necessary, then, to find computationally inexpensive 
and sufficiently accurate estimators of the ionization state of the gas.

The main three quantities that can be used to estimate ionization 
are the above mentioned $J_{\rm{ion}}$, gas number density $n$, 
and mean photon energy, $\E{h\nu} \equiv \int_0^\infty h\nu J_\nu/J\ d\nu$, 
where $h$ is Planck's constant.
In optically thin plasmas, the ionization state is determined 
by the ratio of the first two quantities, as shown by \citet{TTS}, 
referred to as the ionization parameter, $\xi \equiv (4\pi)^2J_{\rm{ion}}/n$.
The mean photon energy is vital for high values of $\xi$, where the main heating 
is Compton heating, which heats the gas to the Compton temperature, $T_C \equiv \E{h\nu}/(4k_B)$.
If $T_C > 10^6$ K, the gas will undergo runaway heating 
and could even become thermally unstable \citep[e.g.,][]{Kallman82, Dyda17, Proga22}. 
We find that, without attenuation of the central source, when the central source has 
just 5\% the luminosity of the disk, all positions above the disk will suffer 
from runaway heating effects, and, even at only 1\%, a significant portion of points 
will overheat as well.

\section{Calculating Radiation and Radiation Force}\label{sec:calc}

To evaluate the properties of the radiation field, it is helpful to start 
with the fundamental equation governing the coupling between radiation 
and matter as a function of position along a line of sight, $\ell$, and time, $t$, 
which is the radiation transfer (RT) equation:
\begin{equation}\label{eq:rt}
    \frac{1}{c}\frac{\partial I_\nu}{\partial t} + \uvec{n}\cdot\nabla I_\nu = j_\nu - \chi_\nu I_\nu,
\end{equation}
where $j_\nu$ is the emissivity and $\chi_\nu$  the total opacity. 
Here, we consider a wind with conditions similar to those studied 
by \citetalias{CAK}. Specifically, we focus  on the time-independent solutions 
and assume that the wind emission is negligible compared to the photospheric emission 
so that we may set $j_\nu$ to zero. In addition, we consider that the opacity due to scattering, 
$\chi^{sc}_\nu$, and absorption, $\chi^{abs}_\nu$,
(i.e., $\chi_\nu=\chi^{sc}_\nu+\chi^{abs}_\nu$ with units $\rm cm^{-1}$) 
for continuum processes. 
This corresponds to the situation where continuum
photons emitted at the photosphere can stream freely through the wind.
Again, following \citetalias{CAK}, we will account for RT in spectral lines 
by using the Sobolev approximation. This approximation  takes advantage 
of the fact that when a photon is emitted at a certain frequency, $\nu$, in the atmosphere,
 it interacts with the outflowing gas through a specific line only when the frequency 
is very close to the line center frequency, $\nu_0$. This resonance point, 
which is determined by the Doppler shift, occurs when the photon frequency 
matches the line center frequency.
Under these assumptions, $I_\nu$ can be treated as constant along any line of sight,
and we need only calculate the intensity of the  photosphere. 
So, for both the corona and the disk,
we let $I_\nu = I_{\nu,0}$ for continuum photons that do not interact via lines,
and $I_\nu = I_{\nu,0}~{\rm exp}(-\tau_\nu)$ for the photons that do interact,
where $\tau_\nu$ is the optical depth due to the absorption line in the wind.

Because $I_{\nu} = I_{\nu,0}$ for continuum photons, calculations of moments 
of intensity can be simplified and, therefore, sped up.
To do this, we use the methods described in \citetalias{Proga99} 
and adopted by \citet{Dyda17}. We use a spherical-polar coordinate system 
$(r_1, \theta_1, \phi_1)$ centered on a point in the wind, $W$
Let the point $W'$ be the point on the disk vertically below $W$.
The coordinate system's zenith is directed along $\overline{WC}$ -- 
where $C$ is the center of the disk -- and $\phi_1$ = 0 
the half-plane opposite $\overline{WC}$ from $W'$.
We also introduce a shorthand $\mu = \cos\theta_1$.
We can write the expressions for $\uvec{n}$ and $\Omega$ in $(r_1, \theta_1, \phi_1)$-space as
 
\begin{equation}\label{eq:nhat def}
    \uvec{n} = (\mu,\ -\sqrt{1-\mu^2}\cos\phi_1,\ \sqrt{1-\mu^2}\sin\phi_1)
\end{equation}
and
\begin{equation}\label{eq:domega def}
    d\Omega = \text{sin}\theta_1\ d\phi_1 d\theta_1 = -\ d\phi_1 d\mu.
\end{equation}

Equations for the zeroth moment, the mean intensity, $J_\nu$, and the first moment, or the flux, $F_\nu$, are quite simple in this coordinate system:

\begin{equation}\label{eq:mean intensity def}
    J_\nu(r,\theta) = \frac{1}{4\pi}\oint_{\Omega(r,\theta)} I_{\nu,0}(r, \theta, \mu, \phi_1)\ d\Omega
\end{equation}

and

\begin{equation}\label{eq:flux def}
    \vec{F}_\nu(r,\theta) = \oint_{\Omega(r,\theta)} I_{\nu,0}(r, \theta, \mu, \phi_1)\uvec{n}(\mu, \phi_1) d\Omega.
\end{equation}
In Equations~(\ref{eq:mean intensity def})~and~(\ref{eq:flux def}), we explicitly show $J_\nu$, $\vec{F}_\nu$, $\Omega$, $I_{\nu,0}$, and $\uvec{n}$ as functions of $r$, $\theta$, $\mu$ and $\phi_1$.
For the sake of brevity, we omit these explicit dependencies in the rest of this paper. 

To compute integrals over d$\Omega$ numerically, we use the Gauss-Legendre quadrature, 
in which we divide the angular $d\mu d\phi_1$-space into a discrete $\Delta\mu \Delta\phi_1$-sum:

\begin{equation}\label{eq:mean_quadrature}
    J_\nu = \frac{1}{4\pi}\sum_i \sum_j I_{\nu,0}(\mu_i, {\phi_1}_j) \omega_i \omega_j
\end{equation}

and

\begin{equation}\label{eq:flux_quadrature}
    \mathbf{F}_\nu = \sum_i \sum_j I_{\nu,0}(\mu_i, {\phi_1}_j)\uvec{n}(\mu_i, {\phi_1}_j) \omega_i \omega_j,
\end{equation}
where $\mu_i$ and ${\phi_1}_j$ are the roots of Legendre polynomials 
while $\omega_i$ and $\omega_j$ are the associated weights 
\citep[for a review of Gauss-Legendre quadrature as well as other methods 
of numerical integration see][]{NR}. In this paper, we used a 15x15 grid of quadrature points, 
which is a relatively low number of grid points, especially compared to older methods 
such as the one used in \citetalias{Proga98}. With such a grid, on a single 
12th Gen Intel i7-12700H Core (i.e., a modern laptop), computing SEDs 
for $J_\nu$ and $\mathbf{F_\nu}$ sampled at 300 frequencies takes 0.1 seconds 
per location, ($r$, $\theta$). Using Gauss-Laguerre Quadrature 
\citep[again, see][]{NR} with 50 frequency quadrature points, we can accurately 
compute $J$ and $\mathbf{F}$ in about 0.02 seconds per location. This computational 
speed makes feasible the goal of incorporating these methods 
in a time-dependent radiation-hydrodynamical simulation.

We obtain an explicit expression for $I_{\nu,0}$ following \citetalias{PSK}'s model 
of radiation from the disk and corona. We separate the intensity into that from each surface: 
the intensity of the disk, $I_{\nu, D}$, and that of the corona, $I_{\nu,*}$.
For the disk, we assume a geometrically thin, optically thick disk with a temperature profile 
given by \citet{Shakura-Sunyaev}. The disk extends from some inner radius, $r_{\rm in}$, 
to some outer radius, $r_{\rm out}$.
We take $r_{in}$ to be $r_{ISCO} = 6GM/c^2$, where $M$ is the mass of the central black hole.
Because the exact shape of the central source is not well known, we assume a spherical extended corona
with a radius also equal to $r_{ISCO}$.
The total luminosity of the disk, $L_D$, is given by 
\begin{equation}
    L_D = \frac{GM\dot{M}}{2r_{in}},
\end{equation}
where $\dot{M}$ is the mass accretion rate. We parametrize the accretion rate in terms 
of the Eddington fraction, $\Gamma \equiv \dot{M}/\dot{M}_{Edd}$, 
where $\dot{M}_{Edd} = 8\pi c r_{in}/\sigma_e$ is the Eddington accretion rate.

We also assume the corona radiates with a power-law spectrum -- i.e. the luminosity $L_{\nu,*} \propto \nu^{\beta}$ 
for some constant $\beta$ -- when $0.1\ keV < h\nu < 511\ keV$ \citep{Sim10}. 
We normalize the corona luminosity by setting, as an input parameter, the ratio $x \equiv L_*/L_D$.
We include further details about the computation of the intensity of the corona in \S\ref{sec:intensity}.
The accretion disk system is parameterized by $M$, $\Gamma$, $r_{\rm out}$, $x$, and $\beta$.
We list the values used in our calculations in Table \ref{tab:disk_params}.

\begin{table}
\begin{center}
\caption{Values of the accretion disk and corona parameters used in this paper.} \label{tab:disk_params}
\begin{tabular}{cc}
\hline
\hline
Parameter & Value(s) \\
\hline
$M$ & $10^8M_\odot$ \\
$\Gamma$ & 0.7 \\
$r_{out}$ & 100 $r_{\rm in}$ \\
$x$ & 0.001-0.25 \\
$\beta$ & -1.1 \\
\hline
\end{tabular}
\end{center}
\end{table}

\subsection{Evaluation of Intensity}\label{sec:intensity}

To develop an expression for the intensity of the disk, we begin with the surface brightness 
of the disk, $Q$. The total surface brightness is the sum of two effects:

\begin{equation}
    Q = Q_{int} + Q_{irr},
\end{equation}
where $Q_{int}$ is the intrinsic brightness of the disk due to internal processes and $Q_{irr}$ is the brightness due to irradiation by an outside source (in our case, the black hole corona). 
The intrinsic brightness is given by the Shakura-Sunyaev model
and the brightness due to irradiation is given by \citetalias{Proga98}:

\begin{equation}
Q_{int} = Q_0\cdot\frac{r_{\rm in}^3}{r_D^3}\left(
    1 - \sqrt{\frac{r_{\rm in}}{r_D}}
\right)
\end{equation}
\begin{equation}
Q_{irr} = \frac{xQ_0}{3\pi}\left[
    \text{arcsin}
        \left(\frac{r_{\rm in}}{r_D}\right)
    - \frac{r_{\rm in}}{r_D}\sqrt{1-\left(
        \frac{r_{\rm in}}{r_D}
    \right)^2}
\right],
\end{equation}
where $Q_0 = 3GM\dot{M}/8\pi r_{\rm in}^3$. 
Assuming that the disk radiates as a blackbody at every point, 
we can use the Stefan-Boltzmann law to write temperature as a function of brightness 
and, thereby, as a function of radius along the disk.

\begin{equation}
    T(r_D) = \left(\frac{Q_{int}(r_D) + Q_{irr}(r_D)}{\sigma}\right)^{1/4},
\end{equation}
where $\sigma$ is the Stefan-Boltzmann constant.

As per our assumption that the disk radiates as a blackbody, we take the intensity 
from a surface element at radius $r_D$ to be the blackbody intensity of that surface element.

\begin{equation}
    I_{\nu,D}(r_D) \equiv B_\nu(T(r_D)),
\end{equation}
where $r_D$ is given by

\begin{align} &
    r_D^2 = r_1^2 + r^2 - 2rr_1 \mu \\&
    r_1 = \frac{r\cos\theta}{\mu\cos\theta - \sqrt{1 - \mu^2}\cos\phi_1\sin\theta}, \nonumber
\end{align}
where $r_1$ is the distance from $W$ to the disk along the line of sight. 
Connecting back to Equations~(\ref{eq:mean intensity def})~and~(\ref{eq:flux def}), 
$I_{\nu, D}$ is a function of $r$, $\theta$, $\mu$ and $\phi_1$ through the fact that $r_D$ 
is a function of those four variables.

We then model the corona's specific luminosity according to a power-law spectrum 
with cutoffs at $h\nu_X = 0.1\ keV$ and $h\nu_\gamma = 511\ keV$.

\begin{equation}
    L_{\nu,*} = \begin{cases}
        xL_{0}\,\nu^{\beta} & \nu_X < \nu < \nu_\gamma\\
        0 & otherwise
    \end{cases}
\end{equation}

where $L_{0}$ is a normalization constant such that $x = L_*/L_D$.
The intensity of the corona is then given in terms of the luminosity,

\begin{equation}\label{eq:istar}
    I_{\nu,*} = \frac{L_{\nu,*}}{4\pi^2r_*^2},
\end{equation}
where $r_*$ is the radius of the corona. For our purposes $r_* = r_{\rm in}$. 

\subsection{Radiation Force due to Lines}\label{sec:line_force}

Computing the radiation force due to a line requires the local line absorption coefficient, 
$\kappa_L$ -- which is related to the opacity in Equation \ref{eq:rt} 
by $\chi_{l,\nu} = \rho\kappa_L\phi(\nu)$, where $\rho$ is the density of the gas 
and $\phi(\nu)$ is the assumed line absorption profile -- and the local intensity, $I_{\nu}$.
\begin{equation}\label{eq:dfrad}
    d\mathbfcal{F}_{L}^{rad} = \frac{\Delta\nu_D\kappa_{L}}{c} I_{\nu}\uvec{n}d\Omega,
\end{equation}
where $\Delta\nu_D \equiv \nu_0v_{th}/c$ is the Doppler width of the line. 
To integrate Equation~(\ref{eq:dfrad}) over d$\Omega$, we cannot simply 
use $I_\nu = I_{\nu,0}$, as this is not a continuum process; 
we must include an escape probability to account for the interaction of photons 
with spectral lines along a line of sight \citep{owocki96}. For a derivation of this escape probability 
from $\phi(\nu)$, see \citet{Lamers-Cassinelli}. With the escape probability included, 
we have an integral for the radiation force due to lines in terms of the intensity of the disk photosphere.

\begin{equation}
    \mathbfcal{F}^{rad}_L = \oint_{\Omega} \frac{\sigma_e\Delta\nu_D}{c} \frac{1-e^{-\tau_L}}{\tau_L}  I_{\nu,0}\uvec{n}d\Omega,
\end{equation}
where $\tau_\nu$ is the local optical depth, which we use the Sobolev approximation 
in three dimensions \citep[see][]{Sobolev} to compute.

\begin{equation}
    \tau_L = \int_{r}^\infty \rho\kappa_L\phi(\nu)\ dr \sobeq \rho\kappa_Lv_{th}\left|\frac{dv_\ell}{d\ell}\right|^{-1},
\end{equation}
where $\phi(\nu)$ is the aforementioned assumed line absorption profile.

To compute the total radiation force due to lines, we follow \citetalias{CAK} 
and define a variable $\eta \equiv \kappa_L/\sigma_e$ so that $\tau_L = \eta t$. 
Substituting in $t$ and $\eta$ into Equation~(\ref{eq:dfrad}), we get an equation 
of radiation force due to a single line:

\begin{equation} \label{eq:Frad L}
    \mathbfcal{F}^{rad}_L = \oint_{\Omega} \frac{\sigma_e\Delta\nu_D}{c}\cdot\frac{1-e^{-\eta t}}{t}\cdot I_{\nu,0}\uvec{n}d\Omega.
\end{equation}
Then, by summing $\mathbfcal{F}^{rad}_L$ over all lines, we obtain the full radiation force due to all lines:
\begin{equation}\label{eq:Frad sum}
    \mathbfcal{F}^{rad} = \sum_{\text{lines}}\oint_{\Omega} \frac{\sigma_e\Delta\nu_D}{c} \frac{1-e^{-\eta t}}{t}\cdot I_{\nu,0}\uvec{n}d\Omega.
\end{equation}

\begin{figure}[H]
    \centering
    \includegraphics[scale=0.35]{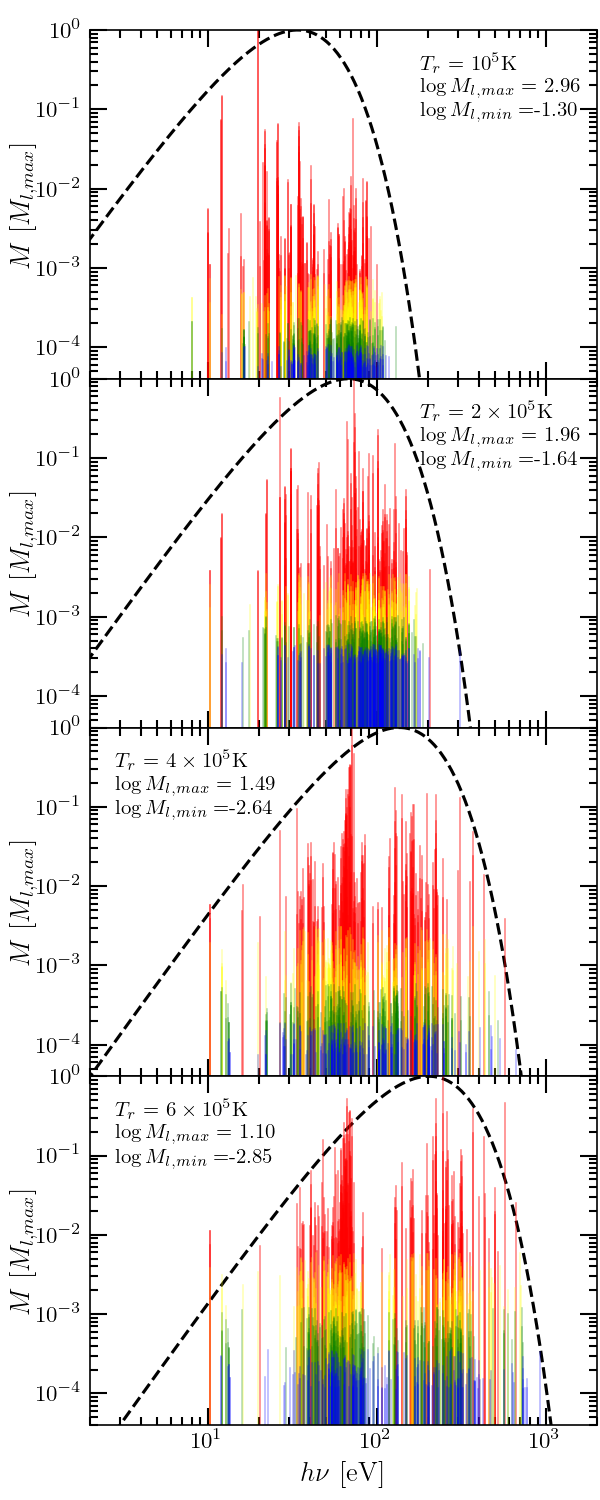}
    \caption{Comparisons of the energy distribution of blackbody radiation (black-dashed curves) and single line force multipliers (colored vertical bars) for four blackbody temperatures: $T_r$ = $10^5$ K, $2\e{5}$ K, $4\e{5}$ K, $6\e{5}$ K.
    The vertical bars represent the 1,000 strongest individual line force multipliers -- normalized to the maximum force multiplier, $M_{l,max}$ -- for fixed values of $\log\xi=2$ and $t=10^{-10}$ (Eq.~\ref{eq:t def}) as a function of energy. These 1,000 strongest lines have been binned into quartiles, red being the upper quartile, yellow the second, green the third, and blue the weakest set of lines.  The black dashed line in each panel represents the assumed blackbody SED (temperature indicated in each panel)  that we used for both determining the ionization balance and computing $M$. The minimum force multiplier, $M_{l,min}$ is also listed in each panel.}
    \label{fig:Line Forest}
\end{figure}

\subsection{The Distribution of Lines}\label{sec:line_dist}

We can use photoionization codes such as \xstar\ \citep{Kallman01} 
to compute the abundances of ions and opacities of lines for any given SED, 
which allows us to compute the sum in Equation~(\ref{eq:Frad sum}).
But, as mentioned before, doing so in the context of radiation-hydrodynamical 
simulations is computationally expensive. 
So, instead, we again follow \citetalias{CAK} and describe 
a statistical distribution of lines, where the number of lines of a given opacity 
and frequency is a function of that opacity and frequency. 
To recover the power law given in \citetalias{CAK}, $M(t) = kt^{-\alpha}$ 
for some constant $k$, we have

\begin{equation}\label{eq:CAK dN}
    dN = N_0\eta^{\alpha-2}\frac{1}{\nu}\ d\eta d\nu,
\end{equation}
where $\alpha$ is the CAK parameter.
\citet[hereafter OCR]{OCR}, add an exponential drop-off to limit the effect 
of very strong driving lines. With this modification, we have
\begin{equation}\label{eq:OCR dN}
    dN = N_0\eta^{\alpha-2}e^{-\eta/\eta_{\rm max}}\frac{1}{\nu} d\eta d\nu,
\end{equation}
where $\eta_{max}$ is the maximum opacity for which lines are physically relevant.

We are interested in finding the distribution of lines relative to energy. 
\citetalias{CAK} introduced a frequency distribution $1/\nu$, which allows 
for the frequency dependency of the force multiplier to cancel out entirely. 
However, based on photoionization calculations using the methods outlined 
by \citet{Dannen19}, we see that the line distribution is not a simple power-law. 
For example, in Figure \ref{fig:Line Forest}, we show that the distribution 
of lines actually traces the energy distribution of the mean radiation intensity. 
This informs our second modification to the line distribution, which is to use 
$J_\nu/J$ to weight our lines by frequency rather than the classic $1/\nu$. 
Combining the modifications of \citetalias{OCR} with our own, we arrive at

\begin{equation}\label{eq:dN}
    dN = N_0\eta^{\alpha-2}e^{-\eta/\eta_{max}}\frac{J_\nu}{J}\ d\eta d\nu,
\end{equation}

We can replace the sum in Equation~(\ref{eq:Frad sum}) with an integral over $dN$ 
and analytically integrate over $d\eta$ to find our, now $\nu$ dependent, force multiplier.

\begin{equation}\label{eq:New M}
    \mathcal{M}(t, \nu) = \frac{\nu J_\nu}{J}\cdot kt^{-\alpha}\cdot\left[\frac{(\tau_{max} + 1)^{1-\alpha} - 1}{\tau_{max}^{1-\alpha}}\right],
\end{equation}
where $\tau_{max} = \eta_{max} t$. Accounting for the SED dependence of the line distribution, 
not only must the force multiplier be integrated over the solid angle of the disk 
but also over all frequencies.  We therefore have
\begin{equation}\label{eq:Frad}
    \mathbfcal{F}^{rad} = \int_0^\infty\oint_{\Omega} \mathcal{M}(t,\nu)\cdot\frac{\sigma_e I_{\nu,0}}{c}\uvec{n} d\Omega d\nu
\end{equation}
as the full expression for the radiation force due to lines. 

\begin{figure}[!htb]
    \centering
    \includegraphics[scale=0.41]{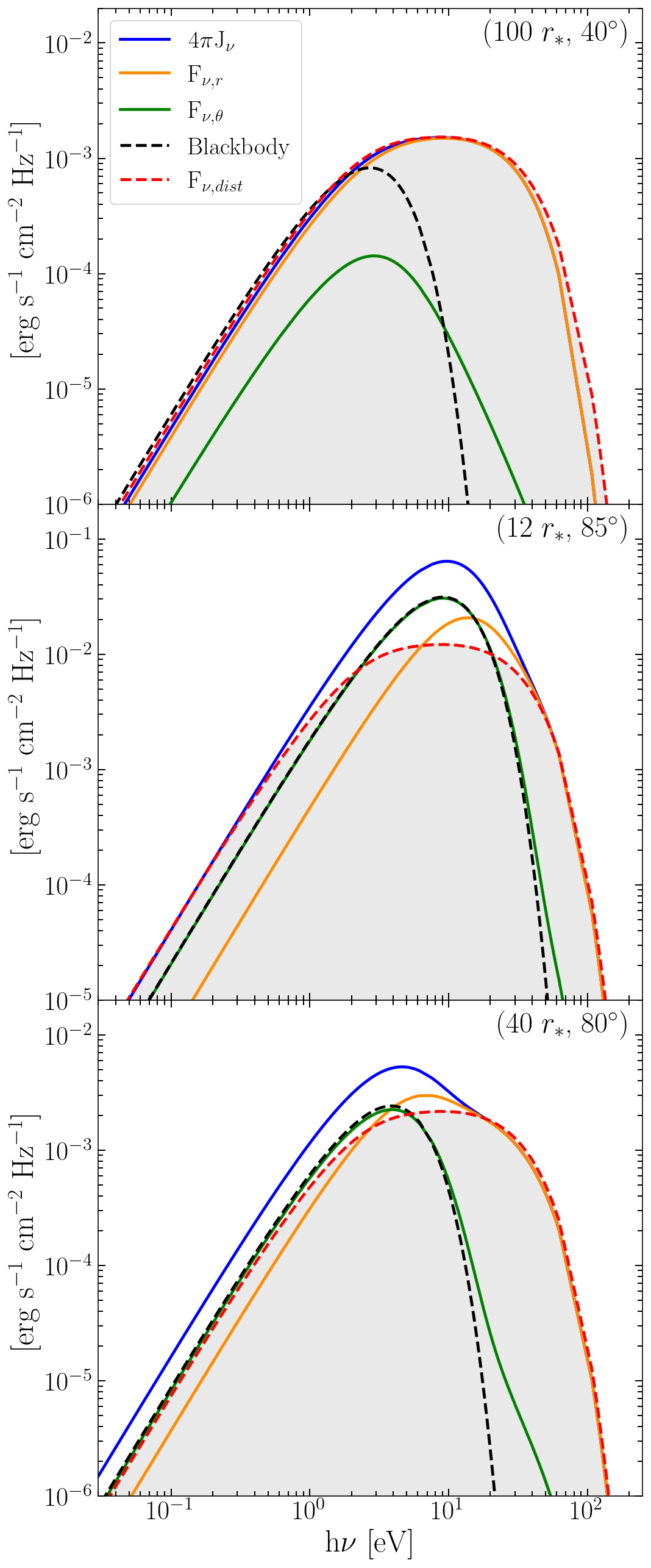}
    \caption{Comparison of the energy distributions of the disk radiation field properties at three locations: (100 $r_{in}$, 40$^\circ$), (12 $r_{in}$, 85$^\circ$), (40 $r_{in}$, 80$^\circ$), top to bottom.
    Each panel shows $4\pi J_\nu$ (blue), $F_{\nu,r}$ (orange), $F_{\nu,\theta}$ (green), $F_{\nu,dist}$ (red dashed), and the blackbody spectrum of the point projected vertically onto the disk (black dashed). The area below $F_{\nu,dist}$ is shaded to make it more apparent at what energies a spectral property is greater or less than that seen by a distant observer.
    }\label{fig:spectra}
\end{figure}
\begin{figure*}[!htb]
    \centering
    \includegraphics[scale=0.27]{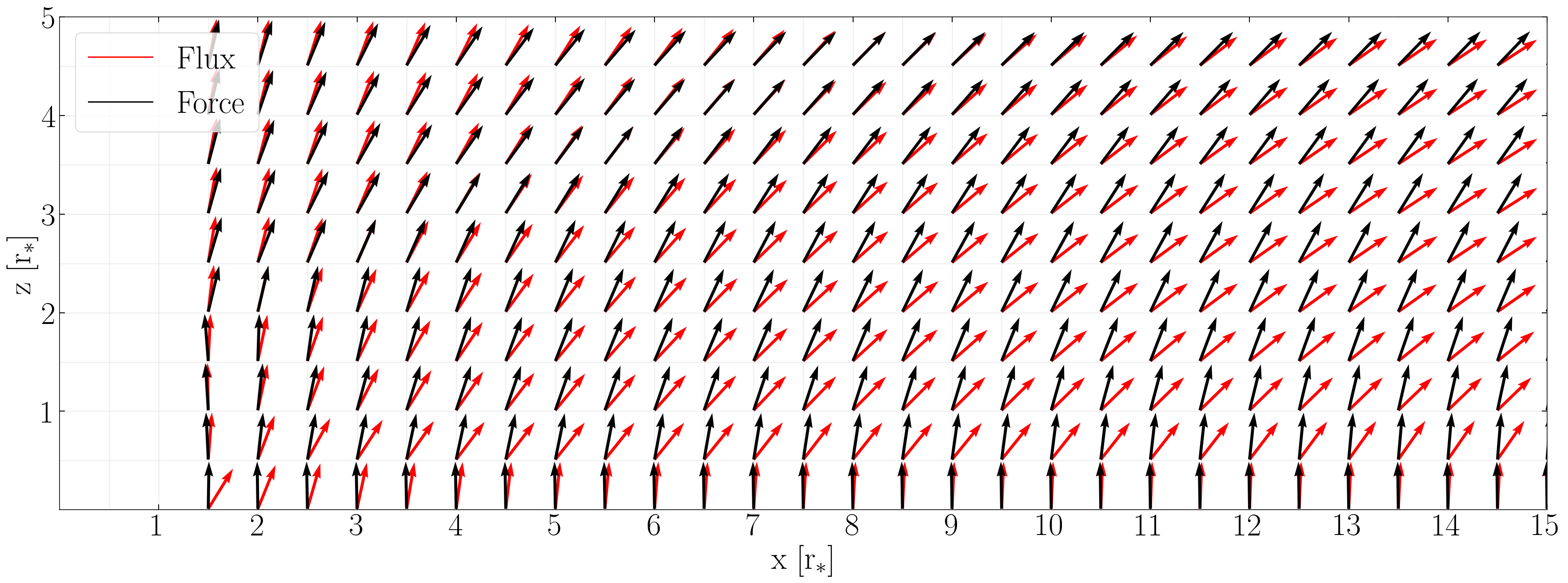}
    \caption{Illustrating the misalignment of the line force and the radiation flux.
    In a vertical velocity field above a disk, the line force (black) is not parallel to
    the radiation flux (red), as it would be in a spherically symmetric wind case.}\label{fig:arrows}
\end{figure*}

\section{Results}\label{sec:res}

\subsection{Local vs Distant Radiation Fields}\label{sec:local vs distant}

We compute energy distributions $J_\nu(r,\theta)$, $F_{\nu,r}(r,\theta)$, and $F_{\nu,\theta}(r,\theta)$ for $r$ ranging from $2r_{in}$ to $100r_{in}$ and
$\theta$ ranging from $0^\circ$ to $90^\circ$.
In Figure \ref{fig:spectra} we show example SEDs for three locations, $(r,\theta)$, above the disk: (100, 40), (40, 80), (12, 85). We selected these three locations as illustrative representations of the three distinct classes of spectra we find above the disk.

Specifically, the top panel of Figure \ref{fig:spectra} shows the spectra at large radius and intermediate theta, ($r=100\ r_{in},\ \theta=40^\circ$).
In this regime, both the spectra of $4\pi J_\nu$ (blue solid line) and $F_{\nu, r}$ (orange solid line) overlap with each other. Moreover, they agree well with the approximation for the flux seen by a distant observer given in Equation~(\ref{eq:F dist}), $F_{\nu, dist}(r,\theta)$ (red dashed line).
Agreement between these three SEDs is consistent with the expectations discussed in \S2.
First, the $\theta$ component of the flux (green solid line) will disagree with $4\pi J_\nu$, $F_{\nu, r}$, and $F_{\nu, dist}$ because it is weighted by $n_\theta$.
And further, $F_{\nu,\theta}$ will be much weaker than $F_{\nu,r}$, which we also see, $F_{\nu,\theta}$ peaks an order of magnitude lower than $F_{\nu,r}$.
For reference, we plot a curve for a blackbody spectrum with the temperature of the point projected down onto the disk.
We see that $F_{\nu,\theta}$ peaks at similar energies to the peak of the blackbody spectrum, but the spectrum is broader, again due to the entire disk contributing to the flux.

In the case that we are very near the disk, shown in the middle panel of Figure \ref{fig:spectra}, ($r=12\ r_{in},\ \theta=85^\circ$),
we see that $F_{\nu,\theta}$ not only peaks at the same energy as the blackbody spectrum, 
but is very similar to it at all energies, save for a small, high-energy tail.
The $J_\nu$ spectrum is the most similar,
in fact, aside from the high energy tail, $4\pi J_\nu = 2F_{\nu,\theta}$.
It is to be expected that contribution to $J_\nu$ is dominated by $F_{\nu, \theta}$ as radiation from any location not immediately below the point $(r, \theta)$ will be diminished by geometric foreshortening.
The other two spectra are quite different from the $F_{\nu,\theta}$, $4\pi J_\nu$, and local blackbody spectra.
While $F_{\nu,r}$ spectrum is relatively narrow, it peaks at higher energies than $J_\nu$ and $F_{\nu,\theta}$ because $F_{\nu,r}$ is weighted by $n_r$, 
and so is dominated by the high energy radiation from the inner disk.
On the other hand, the $F_{\nu,dist}$ spectrum is still a broad multi-temperature blackbody, as it's intensity varies uniformly, independent of frequency.

From these two cases, it may seem reasonable to assume that as one moves 
further from the disk, $J_\nu$ broadens, $F_{\nu,r}$ broadens and increases in power, 
and $F_{\nu,\theta}$ broadens slightly but decreases in power.
However, the bottom panel of Figure~\ref{fig:spectra}, ($40\ r_{in},\ \theta=80^\circ$), 
shows that a simple interpolating function between the narrow spectum near the disk and broad distant observer spectrum does not reproduce the spectra at intermediate positions.
At points with intermediate radii and fairly high viewing angles, 
$F_{\nu,r}$ experiences contributions from a larger portion of the inner disk, 
broadening its spectrum. The part of the disk closer to the point in the wind contributes 
more to the spectrum and is at lower energies. This leads to a low energy bump 
in the $F_{\nu,r}$ spectrum. The $J_\nu$ spectrum has a more prominent bump that covers 
both the $F_{\nu,r}$ and $F_{\nu,\theta}$ peaks.

\begin{figure*}
    \centering
    \hspace{-0.5cm}\includegraphics[scale=0.4]{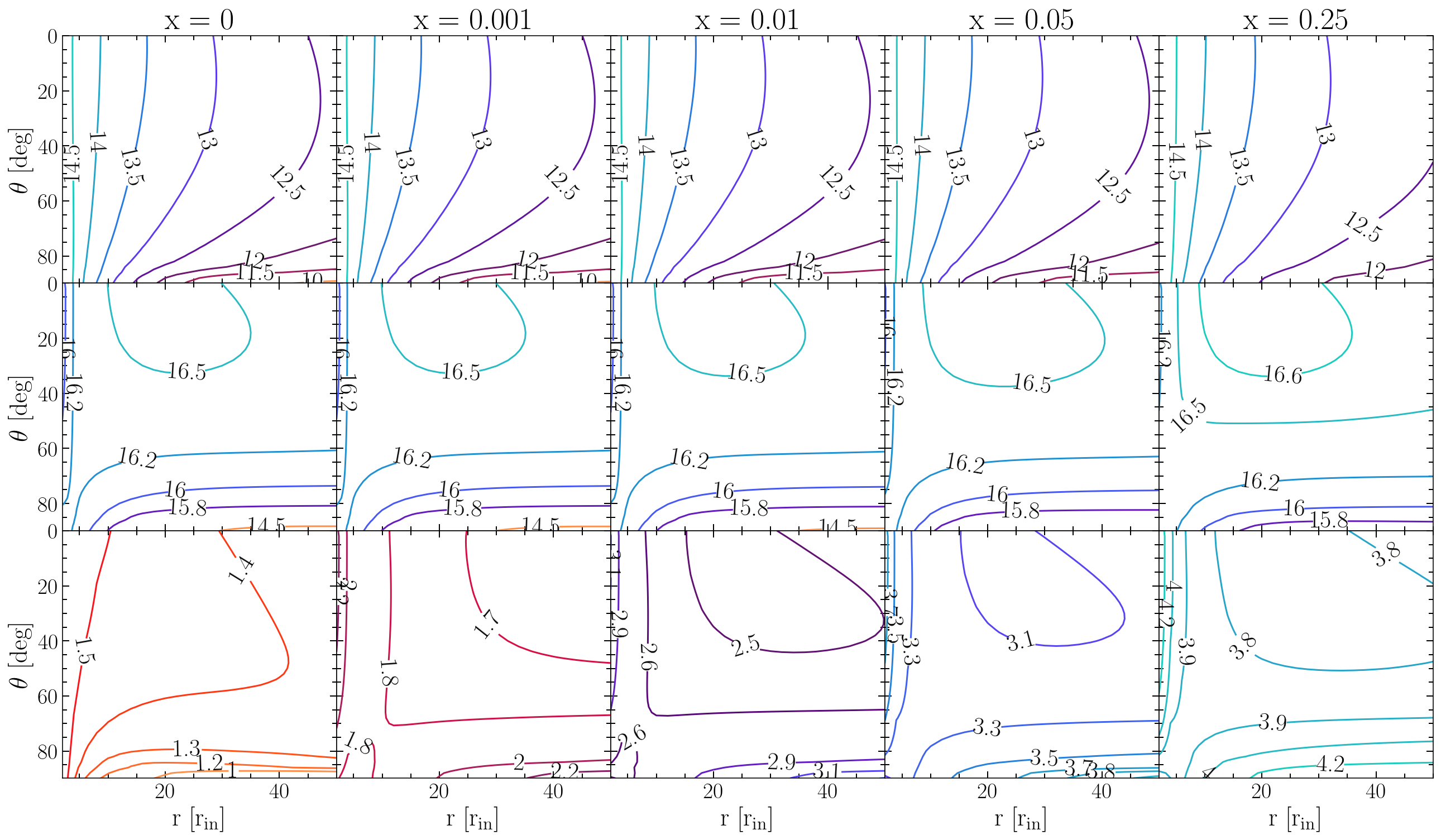}
    \caption{Position dependence of log$_{10}(J_{\rm{ion}})$ and log$_{10}(\E{h\nu})$ for different corona intensities.
    The top two rows of panels show contours of $J_{\rm{ion}}$ in units of erg s$^{-1}$ cm$^{-2}$, both on its own (top) and scaled by $\mathcal{W}(r)$ (middle). 
    The bottom panel shows $\E{h\nu}$ in units of eV.
    Each column is computed for a coronal luminosity fraction $x$.
    The colors are for visual contrast only.
    We demonstrate the position dependence of these quantities 
    as well as the change in that position dependence due to the luminosity of the corona.
    }\label{fig:contour corona comp}
\end{figure*}

\begin{figure*}
    \centering
    \includegraphics[scale=0.47]{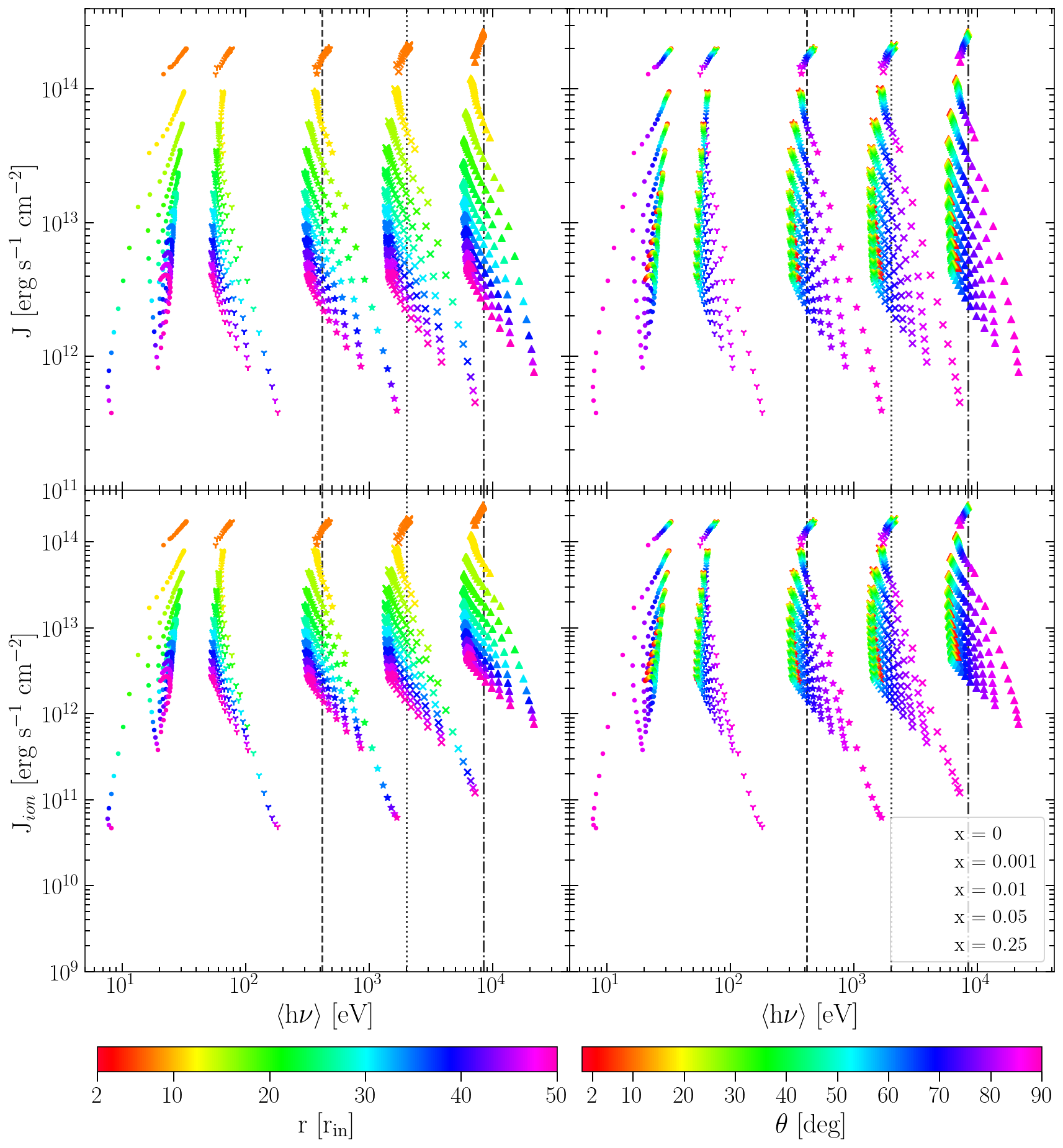}
    \caption{Relations between the energy integrated properties of the disk radiation field with and without coronal radiation.
    The top panels show $\E{h\nu}$ vs $J$ and the bottom show $\E{h\nu}$ vs $J_{\rm{ion}}$.
    The left column shows each point colored according to $r$. The right shows each point colored according to $\theta$.
    The vertical lines are an analytic estimate for $\E{h\nu}$ based on the assumption that $J_*/J_D \approx x$; dashed is $x$ = 0.01, dotted is $x$ = 0.05, and dash-dotted is $x$ = 0.25.
    We do not show these vertical lines for $x = 0$ and $x = 0.001$ because the approximation only holds when $x \gg \E{h\nu}_D/\E{h\nu}_* \sim 10^{-3}$
    for our disk parameters. 
    The maximum value of $\E{h\nu}$ in each panel is $\E{h\nu}_* \approx$ 42 keV.}\label{fig:birds}
\end{figure*}

The position dependences of $J_\nu$ are important for our discussion 
of the radiation force due to lines, as they determine what lines are available 
for line driving, as discussed in \S\ref{sec:line_dist}.
This means that in some sense $\mathbfcal{F}^{rad}$ is doubly position dependent. 
It depends on the position like SEDs do directly through its integration over $\Omega$.
But also, it is position dependent on what lines are available through $J_\nu$.
In the bottom panel of Figure~\ref{fig:spectra}, we see that the $J_\nu$ spectrum 
peaks at a very different position than the $F_{\nu,r}$ spectrum,
but relatively close to the peak of the $F_{\nu,\theta}$ spectrum.
This biases the force toward the $\theta$ direction, as there will be 
more lines that the radiation comprising $F_{\nu,\theta}$ can interact with. 
Such a bias creates a misalignment between the radiation force and the flux.
The velocity of the gas further biases which component of the flux predominantly 
contributes to the line force. In the following subsection, we discuss 
these two effects and determine their relative contribution to causing the misalignment.

\subsection{The Misalignment of $\mathbfcal{F}^{rad}$ and $\vec{F}$}\label{sec:misalign}

For a given velocity field, we can compute $dv_\ell/d\ell$ for each line of sight to the disk, 
and with gas parameters $k$, $\alpha$, $v_{th}$, and $\rho$, we are then able to compute 
the radiation force using Equation~(\ref{eq:Frad}). The radiation force is 
a function of three quantities: the flux from the disk, the velocity gradient 
of the accelerating gas, and the lines available for line driving. 
The weighting effects of the velocity field cause the force to be biased toward 
the direction of greatest acceleration, as it has the greatest value of $kt^{-\alpha}$.
The weighting effects of line availability cause the force to be more biased 
by local radiation than the flux alone would imply. For example, when the point 
in the wind, $W$, is near enough to the disk, local radiation is the dominant contributor 
to $J_\nu$ (see Fig.~\ref{fig:spectra}).
Because the available lines are determined by $J_\nu$, these lines will be primarily 
in frequencies produced by the disk directly below $W$, which creates a vertical bias 
in the force. The biases due to both the velocity field and the local SED cause 
a misalignment between the flux and the force based on the strength of the directional 
force multiplier.

Figure \ref{fig:arrows} shows an example of this misalignment.
The red arrows show the radiation flux purely from the disk ($x=0$).
The black arrows show the radiation force due to this flux in a vertical velocity field, 
which is the velocity field we expect to see very near the disk.
This velocity field is defined by a power-law of the vertical distance. 
In cylindrical coordinates $(r_D, z, \phi_D)$
\begin{equation}
    v_z = v_0\left(\frac{z}{r_D}\right)^\delta,
\end{equation}
where $v_0 = \sqrt{GM/r_D}$, and in spherical coordinates
\begin{equation}
    v_z = \sqrt{\frac{GM}{\sin\theta}}\frac{1}{\tan^\delta\theta}\ r^{-1/2}.
\end{equation} \\
In a Keplerian disk, $v_\phi = v_0$ giving us an expression for $\vec{v}$ in spherical coordinates:
\begin{equation}
    \vec{v} = v_z\left(\cos\theta, -\sin\theta, \tan^\delta\theta\right),
\end{equation}
which we can use to compute $\mathcal{M}(\nu,t)$.

Figure \ref{fig:arrows} shows the misalignment for $\delta = 1/2$.
As one moves vertically upwards from the disk surface, 
the radiation flux is significantly influenced by its radial component.
But, due to the vertical velocity field,
the radiation force remains close to vertical for much longer than the flux.
This allows the gas to potentially be launched higher than if the force were parallel to the flux.

As mentioned, this misalignment comes from a mixture of two effects: the velocity field and the position dependence of the line distribution.
To determine which is the dominant effect,
we computed the directions of the radiation force with the classical $1/\nu$ weighting of frequencies (for which all frequency dependence of the force multiplier cancels).
We found the difference between the misalignment obtained using Equation~(\ref{eq:New M}) and that obtained using the $1/\nu$ weighting.
The points with the largest difference between the two methods differed by about 10\%. 
We conclude that the directionality caused by the velocity gradient is the dominant effect 
causing this misalignment; but the effect of accounting for the frequency-weighting of lines is not insignificant.

\subsection{Position Dependencies of Frequency Integrated Radiation Fields}\label{sec:x_sensitivity}

Along with the radiation force due to lines, other frequency-integrated properties help determine the state of the gas.
The properties of the disk that we focus on in this study are $\E{h\nu}$ and $J_{ion}$. 
These two quantities largely control the heating and ionization of the gas surrounding the disk, as discussed in \S\ref{sec:prelim}.
As spectra change with a position near the disk, so do these frequency-integrated quantities.
In Figure \ref{fig:contour corona comp} we show log$_{10}(J_{\rm ion})$ (top), log$_{10}(J_{\rm ion}/\mathcal{W}(r))$ (middle), and log$_{10}(\E{h\nu})$ (bottom).

In the case where there is no corona ($x=0$), we see a simple trend: as we move radially outward, both $J_{ion}$ and $\E{h\nu}$ decrease.
For large enough $r$ and $\theta$, $J_{ion}/\mathcal{W}(r)$ is radius independent,
consistent with a $\mathcal{W}(r)$ scaling.
Moving toward the disk along constant $r < 20\ r_{in}$, 
$J_{ion}$ and $\E{h\nu}$ both decrease as $\theta$ increases.
At radii larger than 20 $r_{in}$, both increase slightly before decreasing.

When a corona is introduced, the shapes of the contours of constant $J_{ion}$ are mainly unaffected. The values of $J_{ion}$ are increased due to a term proportional to $\mathcal{W}(r)$ added by the corona. This term never varies by more than a factor of $\frac{1}{2}\mathcal{W}(r)$ across all $\theta$, contributing a relatively constant factor at any given radius.

The effect of a corona on $\E{h\nu}$ is much more complicated than on $J_{ion}$. 
Here we introduce a variable $x_J \equiv J_*/J_D$,
where $J_*$ and $J_D$ are the frequency-integrated mean intensities of the corona and disk, respectively.
We rewrite mean photon energy as
\begin{equation}\label{eq:mpe_xJ_decomp}
    \E{h\nu} = \frac{1}{x_J+1}\E{h\nu}_{D} + \frac{x_J}{x_J+1}\E{h\nu}_*,
\end{equation}
where $\E{h\nu}_{D}$ and $\E{h\nu}_*$ are the mean photon energies of the disk and corona, respectively.
This form of the equation is rather useful because $\E{h\nu}_*$ is a constant; 
it is approximately 42 keV for our disk parameters.
As $r$ increases, $x_J$ approaches $x$ and $\E{h\nu}_D$ approaches a constant $\E{h\nu}_{dist}$ -- the mean photon energy of the disk as seen by a distant observer --
which for our disk parameters is $\sim$25 eV.

For constant $\theta \lesssim 40^\circ$, as $r$ increases, first $\E{h\nu}$ decreases. 
This is the regime wherein the radiation from the outer disk becomes more relevant,
softening the mean intensity spectrum. Then, at large $r$, $\E{h\nu}$ begins 
increasing with $r$. As $r$ increases, $x_J$ approaches $x$, and so $\E{h\nu}$ 
approaches it's limit: $x/(x+1)\cdot4.2\e{4} + 1/(x+1)\cdot25$ eV. 
For our four values of $x$: 0.001, 0.01, 0.05, and 0.25, these limits 
are 67~eV, 440~eV, 2~keV, and 8.4 keV, respectively.

The behavior when $\theta \gtrsim 40^\circ$ is the opposite of its behavior in the $x=0$ case.
As one moves closer to the disk while maintaining constant $r$, $\E{h\nu}$ \textit{increases}.
The cause is that as one approaches the disk,
the mean intensity spectrum experienced loses its high energy tail due to the foreshortening effect discussed in \S\ref{sec:local vs distant}, so the total mean intensity of the disk decreases.
The corona does not suffer any foreshortening
and, therefore, becomes the dominant contributor to $\E{h\nu}$.

In Figure~\ref{fig:birds} we show the relation between J, $J_{\rm{ion}}$, and $\E{h\nu}$ more directly. 
We can see that the change in contour shape present in Figure~\ref{fig:contour corona comp} translates to a reversal in the correlation of $J$ and $\E{h\nu}$.
Namely, when there is no corona, 
when one fixes a value of either $r$ or $\theta$ and allows the other to vary,
we find a positive correlation between $J$ and $\E{h\nu}$.
However, when a corona is added, most fixed $r$ and $\theta$ yield negative correlations.
Further, from Equation~(\ref{eq:mpe_xJ_decomp}), if $x_J \gg \E{h\nu}_* / \E{h\nu}_D \sim 10^{-3}$ we can ignore the disk contribution to $\E{h\nu}$ and use $\E{h\nu} = x_J/(x_J+1)\E{h\nu}_*$.
In Figure \ref{fig:birds}, we show $x/(x+1)\E{h\nu}_*$, 
which, if one assumes $x_J \approx x$ as it does at large distances, should be a good estimate of $\E{h\nu}$.
This benchmark is represented by three vertical lines at $x/(x+1)\E{h\nu}_*$ for $x$ = 0.01, 0.05, and 0.25 (dashed, dotted, and dash-dotted lines, respectively).
In all three cases, this benchmark overestimates $\E{h\nu}$ for most points near the disk, indicating $x_J \not\approx x$; rather, $x_J < x$, indicating that near the disk, the coronal contribution to $\E{h\nu}$ is less than one would assume from the disk and coronal luminosities.

\section{Concluding Remarks}\label{sec:conclusions}
We have investigated how the properties of the radiation field 
near an accretion disk change with a position due to purely geometric effects in the case of a standard, flat Shakura-Sunyaev disk surrounded by optically thin gas. 
As explained in \S\ref{sec:prelim}, quantifying this position dependence is a necessary prerequisite for accurately modeling important processes in radiation-hydrodynamical simulations, such as radiative heating and cooling, and the radiation force due to lines.
We described our numerical methods for computing, in a position-dependent manner, various radiation field characteristics, including energy distributions of mean intensity and flux, the line force, the mean ionizing intensity, and the mean photon energy.

Typically, the energy distributions of mean intensity and flux components differ, 
and their position dependence is non-trivial. However, certain distributions 
are relatively straightforward to estimate in two limiting cases.
Firstly, at the locations near the disk, the $\theta$ component 
of the flux is characterized by a single-temperature blackbody distribution.
This flux originates from a small, almost constant temperature disk 
area directly below. 
Secondly, at locations far from the disk, the mean intensity 
and radial flux distributions resemble multi-temperature blackbody spectra 
originating from the entire disk.
Using our methods, we can recover the dependencies in these two regions very well
(e.g., compare the solid green and dashed black curves in the middle 
panel of Fig.~\ref{fig:spectra} and  the solid blue and orange curves to the dashed red in the top
panel of that figure).
In addition, we find that the geometric dilution factor
well captures the radius dependence of the mean intensity for
$r~\gtrsim~15~r_{\rm in}$ and $\theta~\gtrsim~40^\circ$.

We also illustrated how
the direction of the radiation force can be affected 
by the position-dependent mean intensity spectrum. 
In accordance with detailed photoionization calculations (see Fig.~\ref{fig:Line Forest}), 
we made the assumption 
that the line opacity’s energy distribution scales like $J_\nu$ 
and showed that the line force can be more perpendicular 
to the disk than the radiation flux vector (see Fig.~\ref{fig:arrows}).

The wind's thermodynamics and observational properties, 
regardless of its launching mechanism, are determined 
by its ionization and radiative heating and cooling rates. 
The temperature and thermal stability of highly ionized gas 
depend on the mean photon energy, $\E{h\nu}$. Therefore,
we calculated $\E{h\nu}$ for different cases and found 
that the mean intensity is much softer than the radial flux 
at many points near the disk.
This softening reduces the severity of the overionization problem as we discussed in the introduction.
We also found that near the disk, 
solely geometrical factors can cause the value of $\E{h\nu}$ to vary 
by up to a factor of 3 when the central source's X-ray luminosity is zero. 
However, even a small increase  in the X-ray luminosity can 
increase $\E{h\nu}$ by several orders of magnitude. 
We note that this dependence on the central luminosity of the X-rays 
can be reduced by attenuation caused by a bulging disk or an inner disk wind.
We will asses the impact of such intervening gas and the misalignment of the line force and flux vectors. 
by carrying out time-dependent radiation-hydrodynamical simulations 
of disk winds.
To do so, we will incorporate our radiation field methods in a radiation-MHD code like \athena.
Our treatment in this work assumes a geometric optics formalism where radiation propagates along linear rays. We neglect special and general relativistic effects. For example, we do not account for Doppler boosting (both Doppler beaming and the Doppler shift) due to emission from a relativistic disk or compute gravitational deflection of photons. Using the treatment presented in Equation~(4) of \cite{yanfei22} we estimated the percent difference in $J$ and $J_\nu$ between including and not including Doppler boosting for the three spectra shown in Figure \ref{fig:spectra}. For the points $(12 r_*, 85^\circ)$, $(40 r_*, 80^\circ)$, and $(100 r_*, 40^\circ)$, we find the errors in $J$ to be, respectively, 3\%, 0.7\%, and 0.6\%. For $J_\nu$, we found the maximum errors to be 3\%, 0.8\%, 1\%.
These errors become progressively worse as one approaches $r_{isco}$, particularly near $\theta = 90^\circ$, where they can grow as large as 50\%. However at such positions that these differences are large the assumption of linear radiation propagation is violated and so a wholly more detailed relativistic treatment is necessary to accurately describe radiation fields near the black hole.

\vspace{0.5cm}
\hspace{-0.5cm}\textit{Acknowledgments}
\vspace{0.5cm}

Support for this work was provided by the National Aeronautics and Space
Administration under TCAN grant 80NSSC21K0496.
D.P. and T.W. acknowledge support from NASA grant HST-GO- 16196.




\end{document}